# From Small to Large: Clos Network for Scaling All-Optical Switching

Jianmin Lin, Zeshan Chang, Liangjia Zong, Sanjay K. Bose, *Senior Member, IEEE*, Tianhai Chang, Gangxiang Shen, *Senior Member, IEEE/Fellow, OPTICA*

*Abstract*—To cater to the demands of our rapidly growing Internet traffic, backbone networks need high-degree reconfigurable optical add/drop multiplexers (ROADMs) to simultaneously support multiple pairs of bi-directional fibers on each link. However, the traditional ROADM architecture based on the Spanke network is too complex to be directly scaled up to construct high-degree ROADMs. In addition, the widely deployed Spine-Leaf datacenter networks (DCNs) based on electrical switches consume too much power and exhibit high packet latency. Because of these issues, Clos networks are considered as promising alternatives for constructing large-scale ROADMs and all-optical DCNs. In this article, we look at a next-generation Clos-based ROADM architecture and show that it indeed provides better blocking performance with lower element and fiber complexities compared with a traditional Spanke-based ROADM architecture. We also discuss the application of a Clos network in all-optical DCNs to show that it can be used to effectively construct large-scale DCNs with significantly greater flexibility in supporting a variety of multicast services and in combining different network topologies.

## I. INTRODUCTION

WITH the advent of the Fifth Generation (5G) era, and the development of several key enabling technologies, applications such as Mobile Edge Computing (MEC), Artificial Intelligence (AI), Internet of Things (IoT) and Internet of Vehicles (IoV) are becoming increasingly popular. This has not only led to the rapid growth of Internet traffic, but has also fueled an increasing demand for high Quality of Service (QoS) with low latency (e.g., 0.5~1 ms latency [1]), high reliability, low power consumption and ubiquitous service. Supporting these features not only require the backbone/backhaul networks to provide stable, high-capacity pipes, but also require datacenter networks (DCNs) with computing resources which operate with sufficiently high speed and high reliability and have low power consumption.

In the backbone networks, the Dense Wavelength-Division Multiplexing (DWDM) technology has already been extensively employed. To efficiently leverage the DWDM technology, Reconfigurable Optical Add/Drop Multiplexers (ROADMs) are also widely deployed to enable flexible all-optical switching. Meanwhile, with the rapid growth of Internet traffic, more fiber pairs (instead of a single pair of bi-directional fibers) are lit on each link in today's networks. This leads to the requirement of ROADMs with higher fiber degrees even though their nodal degrees may be unchanged. Moreover, it is anticipated that this trend will continue over time with increasing traffic and more fibers being deployed. This therefore raises the important question of *how the traditional ROADM architecture should evolve to support higher fiber degrees to sustain this rapid growth of Internet traffic*.

On the other hand, in data center networks (DCNs), the (folded) Clos (Spine-Leaf) networks have been employed as the switching architecture for decades. However, today's DCNs significantly rely on electrical switches, which leads to several disadvantages, like small capacity, high power consumption, and long latency. To overcome these disadvantages, an optical switching technology may be a promising alternative to replace electrical switches or co-work with them in next-generation DCNs. However, when using all-optical switching technology, an open question for all-optical DCNs will be *whether the Clos switching architecture will remain competitive for these all-optical DCNs*.

This paper tries to answer the above two questions. Specifically, we first elaborate on the traditional ROADM architecture and its associated features. Based on this, we further discuss new ROADM architectures that are being evolved based on the Clos network. We consider various Clos-based ROADMs with different optical switching elements and compare their costs and performance. Finally, we discuss and evaluate the potential of applying the Clos network to optical DCNs.

## II. CLOS NETWORK IN BACKBONE NETWORKS: CLOS-BASED ROADM ARCHITECTURE

### A. ROADM Features

ROADM is a key switching component in today's backbone optical networks. It consists of two main parts, i.e., line side and add/drop side (A/D side) [2]. Each *line side* consists of a pair of

Jiemin Lin and Gangxiang Shen are with the School of Electronic and Information Engineering, Suzhou Key Laboratory of Advanced Optical Communication Network Technology, Jiangsu Engineering Research Center of Novel Optical Fiber Technology and Communication Network, Soochow University, China (Corresponding author and e-mail: Gangxiang Shen, shengx@suda.edu.cn).

Zeshan Chang, Liangjia Zong, and Tianhai Chang are with the Huawei Technology, Dongguan, Guangdong Province, 610212, China.

Sanjay K. Bose is with the Plaksha University, Mohali, 140306, India.

ingress and egress modules. An ingress module distributes optical connections (i.e., wavelengths) to different egress/drop modules, and an egress module aggregates optical connections (i.e., wavelengths) from different ingress/add modules. Each *add/drop side* consists of a pair of add/drop modules. An add module relays optical connections from local terminals to different egress modules, and a drop module distributes optical connections from different ingress modules to different local terminals. Here, each local terminal carries one wavelength.

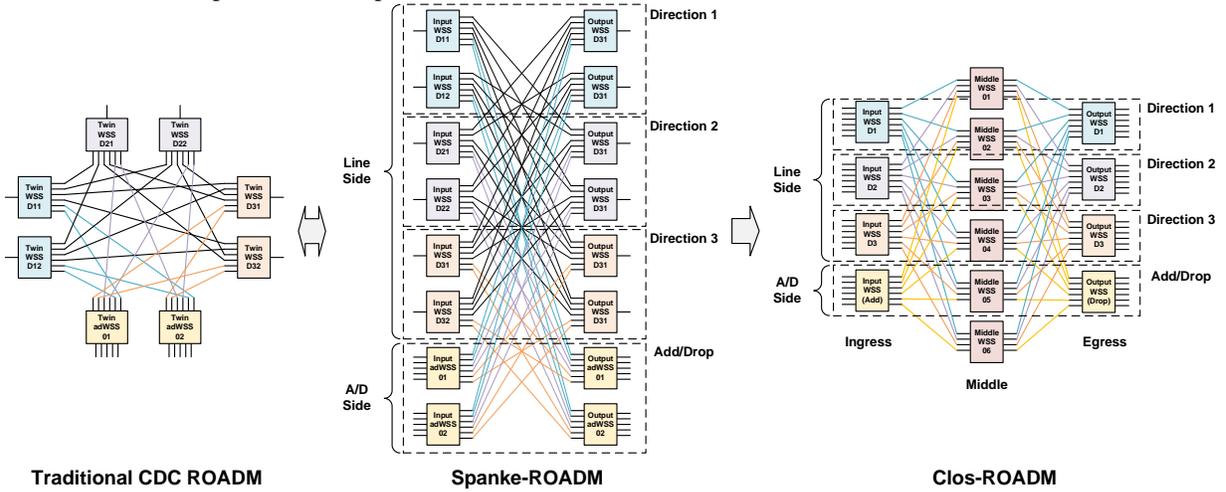

|  | Spanke-ROADM $s(D,L)$ | Clos-ROADM $v(M,D,L)$ |
|---|---|---|
| # of elements | $2 \cdot D \cdot L$ | $2 \cdot (D + L)$ |
| # of fibers | $(D^2 - D) \cdot L^2$ | $4 \cdot L^2$ |

Fig. 1. Architectures and complexities of CDC-ROADM, Spanke-ROADM, and Clos-ROADM.

ROADMs are expected to support the three key features of being *colorless*, *directionless*, and *contentionless*, which are defined as follows [3]. *Colorless* means that each *add/drop* port of a ROADM should not be wavelength-selective, and any wavelength can be added/dropped at an *add/drop* port. *Directionless* means that each *add/drop* port is not nodal degree selective, and any optical connection added on a port can be directed to any egress module, and vice versa. *Contentionless* means that, in a ROADM, establishing optical connections between add/drop ports and ingress/egress modules will not prevent other optical connections from being set up, and that if there is a free add/drop port and a free wavelength on an ingress/egress module, an optical connection can always be set up between them.

*B. Spanke-based ROADMs*

A ROADM supporting the *colorless*, *directionless*, and *contentionless* features is called a CDC ROADM. Fig. 1 shows the basic architecture of a CDC ROADM (see the left-hand side) [2-3], which is made up of switching components, i.e., Wavelength Selective Switches (WSSs). On the line side, $1 \times K$ WSSs and $K \times 1$ WSSs are deployed as the ingress/egress modules, respectively. On the add/drop side, $M \times N$ WSSs are employed to add/drop wavelengths. The ingress/egress modules and add/drop modules are fully connected by short-reach fibers in the backplane of the ROADM. Although a CDC ROADM is often displayed in the format as shown on the left-hand side of Fig. 1, it is essentially a Spanke ROADM as shown in the middle of Fig. 1 if a transformation is made for its backplane [4].

In the new Spanke format, the line side of a ROADM is related to degrees, based on which there are two different types of degrees, i.e., *directional* degree and *fiber degree*. Each directional degree corresponds to a geographic degree of a ROADM node in a network topology, while the fiber degree corresponds to a pair of bi-directional fibers contained on a directional degree. Since there can be multiple pairs of bi-directional fibers on a directional degree, multiple fiber degrees can share a common directional degree. We define $s(D, L)$ as a Spanke-ROADM containing $D$ directional degrees with $L$ fiber degrees on each directional degree.

The Spanke-ROADM can strictly ensure internal non-blocking through its fully-connected backplane. Nonetheless, it would require a huge number of short-reach fibers and will have a low scalability when a large-scale ROADM is constructed. This disadvantage would become even more severe when higher-degree ROADMs are required.

*C. Clos-based ROADM Architecture*

Seventy years ago, C. Clos designed a useful switching network, called a Clos network [5], for the telephone switching network. It allows the use of small-scale (strictly non-blocking) switching elements to construct a large-scale switching network, while still guaranteeing the strictly non-blocking feature. The Clos network consists of three switch stages, i.e., input, middle, and output stages. It owes better scalability when constructing a large-scale switch compared with the Spanke network. However, to construct a Clos network, $M \times N$ switching elements are required. In the past, $M \times N$ WSS technologies were premature and there were no commercial $M \times N$ WSSs. Today, $M \times N$ WSSs are gradually becoming mature [6-7], we can consider replacing $1 \times K$ WSSs using $M \times N$ WSSs for constructing larger-scale ROADMs. The right-hand side of Fig. 1 illustrates a ROADM based on the Clos network (Clos-ROADM). The Clos-ROADM consists of an ingress stage, a middle stage, and an egress stage. Two neighboring stages are interconnected by a fully connected network using short-reach fibers. Switching elements in the ingress and egress stages consist of the line and the A/D sides of the ROADM and switching elements in the middle stage relay the ingress and egress switch stages and provide different routes for connections established between the two stages. For a Clos-ROADM with $D$ directional degrees and $L$ fiber degrees, the ingress and egress stages require arrays of $L \times M$ (or $M \times L$) WSSs and the middle stage requires an array of $D \times D$ WSSs. As in [9], we represent a Clos-ROADM containing $M$ middle-stage switching elements and $D$ directional degrees with $L$ fiber degrees as $v(M, L, D)$.

Recently, there is an increasing interest on how to construct ROADMs based on the Clos network in both academia [8-9] and industry [10-11]. In [8], an initial Clos-ROADM was proposed. In [9], strictly non-blocking conditions for WSS-based Clos-ROADMs were derived, which provides a theoretical foundation for the Clos-ROADMs. In [10-11], top industrial vendors paid special attention to the potential of Clos-ROADM and verified its performance based on simulations.

*D. Performance Comparison between Spanke-ROADM and Clos-ROADM*

We compare Clos-ROADMs with Spanke-ROADMs in terms of their respective element complexity, fiber complexity, and blocking performance.

**Element and Fiber Complexities**: We first consider the aspects of element and fiber complexities. The element complexity is referred to as the number of elements required in a ROADM, and the fiber complexity is referred to as the number of fibers required in a ROADM. In a $s(D, L)$ Spanke-ROADM, $2 \cdot L$ WSSs are required for $L$ pairs of bi-directional fibers in each directional degree, and therefore, the total number of WSSs required is $2 \cdot L \cdot D$ for a Spanke-ROADM with $D$ directional degrees. In contrast, in a $v(M, L, D)$ Clos-ROADM, $2 \cdot D$ WSSs are required in the ingress and egress stages when there are $D$ directional degrees and $M$ WSSs are required in the middle stage. Thus, the total number of WSSs required is $2 \cdot D + M$. Note that here we only count the number of switching elements, but do not consider the difference between $1 \times K$ and $M \times N$ switching elements though the latter can be more expensive than the former.

In a $s(D, L)$ Spanke-ROADM, one fiber degree requires two $1 \times (D - 1) \cdot L$ WSSs for incoming connections and outgoing connections (Note that WSSs on the same direction degree are not inter-connected to each other). Thus, each fiber degree requires $(D - 1) \cdot L$ fibers. Since there are $D \cdot L$ fiber degrees, the total number of fibers required in this architecture is $(D^2 - D) \cdot L^2$. In contrast, in a $v(M, L, D)$ Clos-ROADM, one ingress switching element needs $M$ fibers to connect with the middle stage, and so does one egress switching element. Therefore, the total number of fibers required in this architecture is $2 \cdot L \cdot M$.

For a Clos-ROADM, its element and fiber complexities are both related to $M$, i.e., the number of middle-stage switching elements. Since Spanke-ROADM is strictly non-blocking for each wavelength (i.e., spatially strictly non-blocking), to ensure a fair comparison, we consider a Clos-ROADM that is strictly non-blocking for each wavelength. The spatially strictly non-blocking condition for the Clos-ROADM is $M > 2 \cdot L - 1$, so we take $M = 2 \cdot L$ for the following comparison as a matter of convenience. As shown in the bottom of Fig. 1, the fiber complexity of a Clos-ROADM is $O(L^2)$, while the fiber complexity of a Spanke-ROADM is $O(D^2 \cdot L^2)$. Thus, the fiber complexity of a Spanke-ROADM is much higher ($D^2$ times) than that of a Clos-ROADM. Similarly, the element complexity of a Clos-ROADM is $O(D + L)$, while that of a Spanke-ROADM is $O(D \cdot L)$, which is therefore much higher than the former. In conclusion, a Clos-ROADM demonstrates significantly greater scalability than a Spanke-ROADM for constructing high-degree ROADMs.

**Blocking Performance**: We also compare the blocking performance of the two types of ROADM architecture. As an example, consider a ROADM with 10 directional degrees and 10 fiber degrees, and supporting 5 wavelengths in each fiber. We evaluate connection blocking performance of the ROADMs based on dynamic traffic load (in Erlang). Specifically, under the dynamic traffic load, connection requests arrive following a Poisson distribution and the holding time of each established service connection follows a negative exponential distribution. Service connections are established between any pair of fiber degrees in different directional degrees. A total of $10^6$ arrived connection requests are simulated, and the connection blocking probability is found as the ratio of the total number of blocked connection requests to the total number of arrivals of connection requests. The offered traffic load for the simulation is 2 Erlang per fiber degree.

To construct such a ROADM, we need a $s(10,10)$ Spanke-ROADM or a $v(M, 10, 10)$ Clos-ROADM. Based on Table I, Spanke-ROADM would need 200 1×90 WSSs and 9,000 fibers for this. In comparison, a Clos-ROADM would need 20 $10 \times M$ (or $M \times 10$) WSSs and $M$ 10×10 WSSs and $20 \times M$ fibers. Here, $M$ is a variable that affects the blocking performance of Clos-ROADM. In [10], $M$ is recommended to be in the range of $[L, 1.3 \cdot L]$. In this study, we set $M = L$ as a performance bound, since it corresponds to the condition of a reconfigurable non-blocking Clos-ROADM [8].

Fig. 2 shows the simulation results of the two ROADM architectures, where the dashed line corresponds to the performance of Clos-ROADM and the solid line corresponds to the performance of a Spanke-ROADM. The blue line indicates the blocking probability, corresponding to the left y-axis of the figure, and the red line indicates the number of fibers, corresponding to the right y-axis of the figure. It is noted that, as $M$ increases, the blocking performance of the Clos-ROADM improves rapidly, and when the blocking probability of Clos-ROADM is very close to that of the Spanke-ROADM, $M = 6$, it corresponds 20 $10 \times 6$ WSSs, 6 $10 \times 10$ WSSs, and 120 fibers. This demonstrates that in addition to significantly saving on the WSSs used, the Clos-ROADM provides large savings on the number of fibers needed (more than 98%) compared with a Spanke-ROADM. This supports our observation that using a Clos-ROADM is not only highly desirable because of its much lower element and fiber complexity than a Spanke-ROADM, but also this advantage would become even more significant when a larger ROADM is required to be constructed.

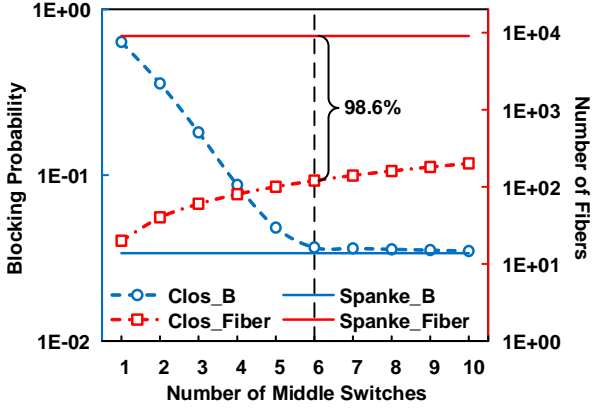

Fig. 2. $s(10,10)$ vs. $v(M,10,10)$.

### III. CLOS-ROADM WITH DIFFERENT MIDDLE-STAGE SWITCHES

The previous section showed the overall performance benefit of a Clos-ROADM over a Spanke-ROADM using WSSs as key switching elements. In this section, we consider other available optical switching elements which may be considered as alternatives for the middle-stage switches of a Clos-ROADM. These may further improve the performance of a Clos-ROADM and also reduce its overall cost.

#### A. Tunable Wavelength Converter-based Clos-ROADM

For simplicity, we will call the Clos-ROADM studied in the previous section as a *WSS Clos-ROADM*. This is subject to the wavelength continuity constraint, which requires that a connection between a pair of input and output ports must use the same wavelength on each passed fiber. The wavelength continuity constraint is critical to a backbone optical network since it is typically topologically sparse. In [9], it is proved that, to satisfy the wavelength continuity constraint and achieve the strictly non-blocking condition, the total number of wavelengths supported by the Clos network needs to be at least twice the number of wavelengths on each line-side port of a ROADM. This therefore poses a great pressure on the WSS element to support many wavelengths as, otherwise, the number of wavelengths supported on each line-side port would become very limited.

To reduce the number of wavelengths required to be supported by a Clos network, we may additionally incorporate wavelength conversion capability in the middle stage of Clos-ROADM as shown in the left top of Fig. 3 (the "TWC-WSS" module). We call this architecture a *TWC-WSS Clos-ROADM*. This is implemented by adding tunable wavelength convertor (TWC) modules at the input ports of the middle-stage WSSs. Here, each TWC module contains a $1 \times K$ de-multiplexer to demultiplex wavelengths, a $1 \times K$ coupler to multiplex wavelengths, and multiple tunable wavelength convertors, each of which can convert wavelengths independently. The detail of these TWCs can also be found in [12]. By introducing these TWC modules, the wavelength continuity constraint in the Clos network can be fully relaxed. Moreover, if an optical network is deployed with this type of ROADM, the wavelength continuity constraint itself can be relaxed in the network. This would significantly improve wavelength assignment flexibility and spectrum resource utilization in the overall network.

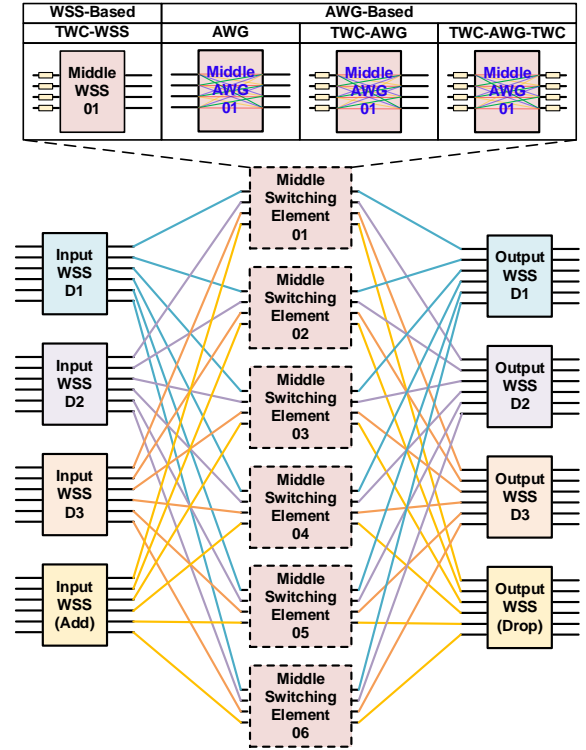

Fig. 3. Clos-ROADMs with different middle switching elements.

#### B. Arrayed Wavelength Gating-based Clos-ROADM

Since WSSs are expensive, a Clos-ROADM with many WSSs would have a high system cost. To reduce the system cost while maintaining switching flexibility, we may employ low-cost switching elements to implement the middle stage of the Clos-ROADM. An Arrayed Wavelength Gating (AWG) may be a good candidate for this as it has a good potential to fulfill the middle-stage switching functionality at lower cost while guaranteeing flexible wavelength-routing capability. The right top of Fig. 3 shows three potential modules for a Clos-ROADM with AWGs and TWCs deployed in the middle stage modules.

The "AWG" module in Fig. 3 uses AWGs to form the middle switching stage (called *AWG Clos-ROADM*). However, this architecture cannot achieve a blocking performance close to WSS Clos-ROADM since AWGs are passive and are not able to switch wavelengths on demand. To improve the wavelength switching flexibility of the middle stage, another option is to use AWGs with the TWC modules. This is expected to improve the blocking performance since TWC modules can change wavelengths. The "TWC-AWG" module in Fig. 3 adds TWC modules before the input ports of the AWGs; we call this a *TWC-AWG Clos-ROADM*. The "TWC-AWG-TWC" module in Fig. 3 further adds TWC modules after the output ports of the AWGs and are called a *TWC-AWG-TWC Clos-ROADM*. Comparing these three AWG-based architectures, there is a tradeoff between system cost and wavelength-switching flexibility. An AWG Clos-ROADM is the cheapest, but the least flexible, while a TWC-AWG-TWC Clos-ROADM is the most flexible, but also the most expensive.

*C. Blocking Performance*

We evaluate the blocking performance of the proposed Clos-ROADM architectures, including the two WSS-based Clos-ROADM (i.e., WSS Clos-ROADM and TWC-WSS Clos-ROADM) and three AWG-based Clos-ROADM (i.e., AWG Clos-ROADM and TWC-AWG Clos-ROADM and TWC-AWG-TWC Clos-ROADM) architectures. The simulation assumptions are as follows. We use $v(5,5,5)$ as the basic Clos-ROADM architecture. Clos-ROADM supports 5 wavelengths in each fiber. Offered traffic load follows the Erlang assumption, i.e., the connection request arrivals between each pair of input-output fibers follows a Poisson process and the holding time of each established connection follows a negative exponential distribution. The offered traffic load between any input-output port pair is the same.

In addition, two special scenarios are considered as benchmarks. One is the traditional CDC-ROADM (Spanke-ROADM), which achieves the best performance among today's ROADMs. The other is a *theoretical limit*, which is calculated based on the following formulae.

$$E_B(\rho, w) = (\rho^w/w!)/(\sum_{k=0}^{w} \rho^k/k!) \quad (1)$$

$$\begin{cases} B_i = E_B(\rho, w) \\ B_o = E_B(\rho(1-B_i), w) \end{cases} \quad (2)$$

$$B = 1 - (1-B_i)(1-B_o) \quad (3)$$

Here, since the incoming traffic is assumed to follow the Erlang distribution, we use the well-known Erlang-B formula (1) to calculate the blocking probability with traffic load $\rho$ and number of available wavelengths $w$. The best performance that a ROADM can achieve is when connection blocking is only due to the lack of free ports, but not due to the internal blocking of the ROADM switching fabric. For this, we can use (2) to calculate the blocking probabilities of the input and output ports ($B_i$ and $B_O$), and finally find the theoretical blocking probability limit of the ROADM using (3).

Fig. 4 shows the blocking performance of the two WSS-based Clos-ROADMs (see blue lines). It is noted that the WSS Clos-ROADM can achieve the same blocking performance as the traditional CDC-ROADM, while TWC-WSS Clos-ROADM can reach the full theoretical limit of blocking performance. This is achieved because of the additional flexibility provided by the TWC modules. Fig. 4 also shows the blocking performance of AWG-based Clos-ROADMs (see the black lines). It is noted that the blocking performance of the AWG Clos-ROADM is not as good as that of the WSS Clos-ROADM. This is because AWGs are inherently less flexible than WSSs. Moreover, the performance improvement by adding TWCs only at the input stage of AWGs is fairly small. This is still attributed to the bottleneck of the middle-stage AWG. However, the blocking performance of AWG-based Clos-ROADM can be significantly improved when TWC modules are added at both the input and output ports of the AWGs. This is because the configuration of TWC-AWG-TWC is essentially the same as TWC-WSS with a full wavelength conversion capability. With this, the TWC-AWG-TWC Clos-ROADM can approach the theoretical limit of blocking performance.

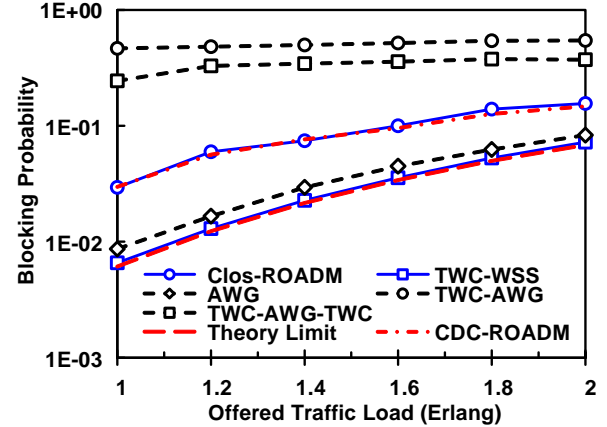

Fig. 4. Blocking performance of WSS and AWG-based Clos-ROADMs.

## IV. CLOS NETWORK IN DATACENTER NETWORKS: FOLDED CLOS ARCHITECTURE

Today's DCNs are typically constructed based on Spine-Leaf networks and using large-scale electrical switches [13]. The disadvantages of these architectures have been widely discussed, mainly including high-power consumption and latency due to electric switching. Optical switching is considered promising to resolve the above issues and is being gradually implemented in DCNs.

*A. Clos Network in All-Optical DCNs*

Spine-Leaf network has many advantages, including a small network diameter and a fixed number of route hops. These advantages are important for all-optical DCNs because the quality of the optical signals can be accurately estimated in this type of network. Thus, the Spine-Leaf network is often employed as a good choice for all-optical DCNs.

A Spine-Leaf network includes a Spine layer and a Leaf layer (see the left-hand side of Fig. 5), which is essentially a (folded) Clos network. The Clos network is an excellent candidate for building a large-scale optical switch using small-scale optical switching elements [14]. The Leaf layer, which corresponds to the stacked ingress and egress stages in a Clos network, interconnects the network devices. The Spine layer provides multiple routes to the Leaf layer, which corresponds exactly to the middle stage of the Clos network. The only difference between Clos and Spine-Leaf networks is the scale of switching elements in the Leaf layer. Specifically, to fold a $v(M,L,D)$ Clos network to a Spine-Leaf network, the size of each switching element in the Leaf layer should be increased from $L \times M$ to $(L+M) \times (L+M)$, since each switching element in the Leaf layer then has $L$ additional input and output ports.

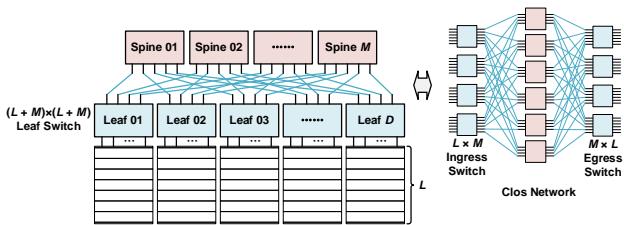

Fig. 5. Spine-Leaf networks.

*B. Variances of Clos Network*

There are two typical types of services in DCNs, i.e., unicast and multicast services. For example, publish-subscribe services for data dissemination are typical multicast services. An optical switch-based network is good at provisioning unicast services, which is however not efficient for multicast services since multiple wavelengths are required for each multicast service. To tackle this issue, we consider employing optical splitters (the diamond module in Fig. 6) to replace spine switches in the Spine layer. Since an optical splitter is passive in equally splitting an optical signal to all output ports, this enables an all-optical Spine-Leaf network with splitters to support multicast services.

Another variance of the Spine-Leaf network is to combine with other topologies, e.g., Torus topology, to efficiently support different types of services, e.g., general datacenter services and High-Performance Computing (HPC) services, in a common DCN. Many HPC systems employ Mesh/Torus topologies because of their high scalability and high performance-to-cost ratio [15]. Fig. 6 shows a 1-D Torus scenario, which replaces a Spine switch with direct fiber connections (the red curves in Fig. 6) to form a Torus topology. By transforming the Spine-Leaf network to an unfolded Clos network, it is interesting to see that the Torus topology essentially employs a round-robin direct connection pattern to replace a middle switch.

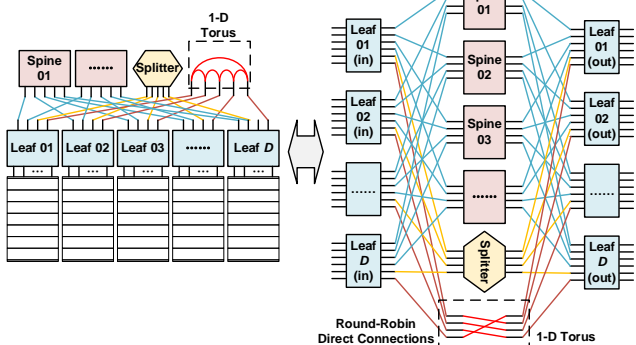

Fig. 6. Spine-Leaf networks with splitters and round-robin direct connections.

## V. CONCLUSION

Increasing traffic demands require large-scale ROADMs, for which the Clos network is considered promising. In this article, we first compared the traditional Spanke-based ROADM and the Clos-based ROADM from the perspectives of element and fiber complexities. Based on the Clos-based ROADM, we further discuss other architectures for improving the blocking performance and reducing the system cost. We demonstrate the tradeoff between blocking performance and system cost for these architectures through simulations. Finally, we also discuss the application of Clos network in all-optical datacenter networks using the Spine-Leaf architecture. Several variations to the basic architecture for supporting different types of datacenter services are also presented.

## BIOGRAPHIES

**Jiemin Lin** (Student Member, IEEE) received B.S. degree from Applied Technology College of Soochow University, Suzhou, China, where he is currently working toward the M.S. degree. His research interests include the area of optical switching and optical datacenter networking.

**Zeshan Chang** received the Ph.D. degree from the City University of Hong Kong, Hong Kong, in 2018. He is currently a Principal Engineer with Huawei Technologies Co., Ltd. His research interest includes the optical cross connects and optical transport networks.

**Liangjia Zong** received the B.Sc. degree in optical information science and technology and the Ph.D. degree in physical electronics from the Huazhong University of Science and Technology, China. His research interests include high speed optical transmission system and all optical switching technology. He joined Optical Research Department, Huawei, as a Research Engineer. He has authored more than 80 papers and patents.

**Sanjay K. Bose** (Senior Member, IEEE) received the B.Tech. degree from IIT Kanpur, Kanpur, in 1976, and the M.S. and Ph.D. degrees from the State University of New York at Stony Brook, Stony Brook, in 1977 and 1980, respectively. From 1980 to 1982, he was with the Corporate Research and Development Center, General Electric Company, Schenectady, NY, USA, with a focus on projects associated with power line communications, optical fiber communications, and mobile-satellite communications. He subsequently joined the faculty of the Department of Electrical Engineering, IIT Kanpur, from 1982 to 2003. From 2003 to 2008, he was on the faculty of the School of EEE, Nanyang Technological University, Singapore. He returned to India, in 2009, and joined the faculty of the Department of EEE, IIT Guwahati, where he was also the Dean, Alumni Affairs, and External Relations, from 2011 to 2014. He is currently a professor in Plaksha University, Mohali, India. He has also held short-term and long-term visiting appointments at the University of Adelaide, the Queensland University of Technology, Nanyang Technological University, and the University of Pretoria.

**Tianhai Chang** graduated from Dalian University of Technology, and jointed Huawei in 2000. He currently is Chief Scientist in Access network domain in Huawei Technology. His research area includes all optical network structure and solution about access network, FTTX, FTTH, FTTR, and indoor wireless.

**Gangxiang Shen** (Senior Member, IEEE) is a Distinguished Professor with the School of Electronic and Information Engineering, Soochow University, Suzhou, China. His research interests include spectrum efficient optical networks, green optical networks, and integrated optical and wireless networks. He has authored and coauthored more than 200 peer-reviewed technical papers. He is currently an Associate Editor for the IEEE/OSA JOURNAL OF LIGHTWAVE TECHNOLOGY, IEEE NETWORKING LETTERS, and an Editorial Board Member of *Optical Switching and Networking* and *Photonic Network Communications*. He was selected as Highly Cited Chinese Researcher by Elsevier in 2014 and 2021. From 2018 to 2019, he was an elected IEEE ComSoc Distinguished Lecturer. From 2018 to 2021, he was a Member at Large of IEEE ComSoc Strategic Planning Standing Committee. He served as a TPC Member for both OFC and ECOC. He is also an elected Fellow of OPTICA.